\documentclass{aa}
\usepackage{graphicx}
\usepackage{natbib}
\usepackage{txfonts}
\begin{document}

\title {An inhomogeneous model for the Galactic halo: a possible explanation 
for the spread observed in s- and r-process elements}

\author {G. Cescutti
\thanks {email to: cescutti@oats.inaf.it}}

\institute{ Dipartimento di Astronomia, Universit\'a di Trieste, via G.B. Tiepolo 11, I-34131}
\date{Received xxxx / Accepted xxxx}

\abstract
{}{ We propose an explanation for the considerable scatter 
of the abundances of neutron capture elements observed in low-metallicity stars 
 in the solar vicinity, compared to the small star-to-star scatter 
observed for the $\alpha$-elements.} {We have developed a stochastic
 chemical evolution model in which the main assumption is a random formation
 of new stars subject to the condition that the cumulative mass distribution
 follows a given initial mass function.} 
{ With our model, we are able to reproduce the different spreads of neutron 
capture elements and $\alpha$-elements in low-metallicity stars.}
{ The reason  for different observed spread in neutron 
capture elements and $\alpha$-elements  resides in the random birth of stars, coupled 
with different stellar mass ranges, from which $\alpha$-elements and neutron capture 
elements originate. In particular, the site of production of $\alpha$-elements is the 
whole range of massive stars, from 10 to 80$M_{\odot}$
 whereas the mass range of production for neutron capture
 elements lies between 12 and 30$M_{\odot}$. }


\titlerunning{Inhomogeneous model for the Galactic halo}

\maketitle

\authorrunning{Cescutti}

\section{Introduction}


Early work by Gilroy et al. (1988) first proposed that the stellar
abundances of very heavy elements with respect to iron, particularly
[Eu/Fe], show a large scatter at low metallicities. 
Their work suggested that this scatter appears to diminish with increasing
metallicity. This was confirmed by the 
 large spread observed later on in the [Ba/Fe] and [Eu/Fe] ratios in halo 
stars (e.g., McWilliam et al. 1995; Ryan et al. 1996).
A more extensive study by Burris et al.
(2000) confirmed the very large star-to-star scatter in the early
Galaxy, while studies of stars at higher metallicities, involving
mostly disk stars (Edvardsson et al. 1993; Woolf, Tomkin, \&
Lambert 1995), show little scatter.
In the last few years, a great deal of observational work on galactic stars 
appeared: Fulbright (2000); Mashonkina \& Gehren (2000, 2001);
Koch \& Edvardsson (2002); Honda et al. (2004); and Ishimaru et al. (2004).
All these works confirmed the presence of the spread for these elements and this
spread can reach 2 dex at [Fe/H]\footnote{We adopt the usual spectroscopic notations that
[A/B]= $log_{10}(N_{A}/N_{B})_{\star}-log_{10}(N_{A}/N_{B})_{\odot}$ 
and that $log\epsilon(A)=log_{10}(N_{A}/N_{H})+12.0$, for elements A and B} $\sim$-3.
It is worth noting that such a high spread is not found for the  
[$\alpha$/Fe] ratios in very metal-poor stars (down to $[Fe/H]=-4.0$, 
Cayrel et al. 2004); in fact, the spread for these elements is around 
0.5 dex at low metallicities.
This fact suggests that the spread, if real, is a characteristic of  these heavy elements
and not only due to an inhomogeneous mixing in the early halo phases,
as suggested by several authors (Tsujimoto et al. 1999; Ishimaru \& Wanajo 1999).

Recently, several studies have attempted to follow the
enrichment history of the Galactic halo with special emphasis given
to the gas dynamical processes occurring in the early
Galaxy: Tsujimoto, Shigeyama, \& Yoshii (1999) provided
an explanation for the spread of Eu observed in the oldest
halo stars in the context of a model of supernova-induced
star formation; Ikuta \& Arimoto (1999) and McWilliam \&
Searle (1999) studied the metal enrichment of the Galactic
halo with the help of a stochastic model aimed at reproducing
the observed Sr abundances; Raiteri et al. (1999)
followed the Galactic evolution of Ba by means of a hydrodynamical
N-body/smoothed particle hydrodynamics code; and Argast et al. (2000)
concentrated on the effects of local inhomogeneities in the
Galactic halo produced by individual supernova (SN) events,
accounting in this way for the observed scatter of some (but
not all) elements typically produced by type II SNe. 
Besides a spread for r-process elements, they also predicted a spread 
of more than 2 dex for  Mg and O at [Fe/H]=-3, which is too large compared 
to the observational data.
Finally, Travaglio et al. (2004) investigated whether incomplete mixing of
gas in the Galactic halo can lead to local chemical inhomogeneities
in the interstellar medium (ISM) involving heavy elements, in particular Eu, Ba, and Sr. 
They reproduced the spread for these elements, but 
left unexplained why the spread is present only
for neutron capture elements whereas it is much smaller for the other 
elements (for example $\alpha$-elements).
 Concerning the $\alpha$-elements, Argast et al. (2002) and Karlsson
 \& Gustafsson (2005) attempted to explain the small dispersion 
in these elements from two different points of view: a different iron 
yield distribution as a function of progenitor mass for Argast et al. (2002),
and  cosmic selection effects favoring contributions from supernovae in a certain mass range 
 for Karlsson \&  Gustafsson (2005).

The purpose of this paper is to explain the different spread observed in 
the halo stars for different elements.
Cescutti et al. (2006) have suggested the best  stellar yields to 
fit the trend of the mean abundances for Eu and Ba in the framework of a homogeneous chemical
evolution model (Chiappini et al. 1997, 2001).
In the present paper, using the same yields of Cescutti et al. (2006), we show
the results of a stochastic chemical evolution model  that we have developed 
to explain the different observed spreads for different
elements in the halo stars. The paper is organized as follows: in Sect. 2 we present the observational 
data; in Sect. 3 the inhomogeneous chemical evolution model is presented; and in Sect. 
4 the adopted nucleosynthesis prescriptions  are described.
In Sect. 5  we present the results;  and in Sect. 6 some conclusions are drawn.

\section{Observational data}
For the extremely  metal-poor stars (-4$<$[Fe/H]$<$ -3), we adopted the recent
results from UVES Large Program "First Star'' (Cayrel  et al. 2004, Fran\c cois et al. 2007).
For the abundances in the remaining range of [Fe/H], we took  high quality data 
 of various sources published in the literature.
These are are sumarized in Table \ref{authors}. 
We homogenized all of these data by normalizing them to the solar abundances by Asplund et al. (2005).

\begin{table*}

\centering
\caption{The authors from which we adopt the observational data for each considered element.}
 \label{authors}

\begin{minipage}{90mm}

\begin{tabular}{|c|c|c|c|c|c|c|c|c|c|}
\hline\hline

	                          & Eu& Ba& La& Sr& Y & Zr& Ca& Mg& Si\\
\hline
Burris et al. (2000)              & X & X &   & X & X & X &   &   &   \\
\hline
Carney et al. (1997)              &   &   &   &   &   &   & X & X & X \\
\hline
Carretta et al. (2002)            &   &   &   &   &   &   & X & X & X \\
\hline
Cayrel  et al. (2004)             &   &   &   &   &   &   & X & X & X \\
\hline
Cowan et al. (2005)               & X &   & X &   &   & X &   &   &   \\
\hline
Edvardsson et al. (1993)          &   & X &   &   & X & X & X & X & X \\
\hline
Fran\c cois et al. (2007)         & X & X & X & X & X & X &   &   &   \\
\hline
Fulbright (2000)                  & X & X &   &   & X & X & X & X & X \\
\hline
Fulbright (2002)		  & X & X &   &   & X & X & X & X & X \\	
\hline
Gilroy et al. (1998)		  & X & X & X & X & X & X &   &   &   \\
\hline
Gratton \& Sneden (1988)	  &   & X &   & X & X &   &   &   &   \\
\hline
Gratton \& Sneden (1994)	  & X & X & X & X & X & X &   &   &   \\
\hline
Honda et al. (2004) 		  & X & X & X & X & X & X & X & X & X \\
\hline
Ishimaru et al. (2004)		  & X & X &   & X &   &   &   &   &   \\
\hline
Johnson (2002)                    & X & X & X & X & X & X & X & X & X \\
\hline
Koch \& Edvardsson (2002)	  & X &   &   &   &   &   &   &   &   \\
\hline
Mashonkina \& Gehren (2000, 2001) & X & X &   & X &   &   &   &   &   \\
\hline
McWilliam et al. (1995)		  & X & X & X & X & X & X & X & X & X \\
\hline
McWilliam \& Rich (1994)	  & X & X & X &   & X & X & X & X & X \\
\hline
Nissen \& Schuster (1997)         &   & X &   &   & X &   &   &   &   \\
\hline
Pompeia et al. (2003)		  &   &   & X &   &   &   &   &   &   \\
\hline
Prochaska et al. (2000)		  & X & X &   &   & X &   & X & X & X \\
\hline
Ryan et al. (1991)		  & X & X &   & X & X &   & X & X & X \\
\hline
Ryan et al. (1996)		  & X & X &   & X & X & X & X & X & X \\
\hline
Stephens (1999)  		  &   & X &   &   & X &   & X & X & X \\
\hline
Stephens \& Boesgaard (2002)      &   & X &   &   & X &   & X & X & X \\

\hline\hline

\end{tabular}

\end{minipage}

\end{table*}

\section{Inhomogeneous chemical evolution model for the Milky Way halo}\label{model}

For comparison with a homogeneous chemical evolution model,
we choose to use the same parameters of the chemical evolution (star formation rate, 
initial mass function, stellar lifetime, nucleosynthesis, gas density threshold) as those 
of the homogeneous model used in Cescutti et al. (2006).
For this reason, we model the chemical evolution of the halo of the Milky Way  for the duration of 1 Gyr.
We consider that the halo consists of many independent cubic regions, each 
with the same typical volume, and each region does not interact with the others. 
 We decided to have a typical volume of $2.8\cdot10^{6}pc^{3}$, in this way
the surface of the volume, taken as the side of a cube, is $2\cdot10^{4}pc^{2}$.
The number of assumed volumes is 100. In order to ensure good statistical
results, we test that a larger number of volumes  and foud that the statistics
 do not  significantly change, but the model is slowed down.
The dimension of the volume is  large enough to allow us to neglect the interaction
among different volumes, at least as a first approximation.
In fact, for typical ISM densities, a supernova remnant becomes 
indistinguishable from the ISM -- i.e., merges with the ISM -- before
reaching $\sim 50pc$  (Thornton et al. 1998), 
less than the size of our cubic region. We do not use larger
volumes because  we would lose the stochasticity
we are looking for; in fact, as we tested for 
increasing volumes, the model tends to be homogeneous.
We use timesteps of 1Myr, which is an interval of time
shorter than any stellar lifetime considered here; in fact, the maximum stellar mass 
considered here is 80$M_{\odot}$ with a lifetime of $\sim$3Myr.
At the same time, this timestep is longer than the cooling time of the SN bubbles, which is 
normally $\sim$ 0.1-0.2Myr and at maximum 0.8 Myr. We describe later
in this same Sect. how we calculate this cooling time. 
With this timestep we can say, in first approximation, that an homogenous mixing 
takes place and, at least for the first period, only a few stars pollute the ISM; this
is again important from the point of view of the stochasticity of the model.

In each region, we assume the same law for the infall of the gas with primordial
 composition; originally, in the homogeneous model,
 we have this law for the halo phase:
\begin{equation}
\frac{d\sigma_{tot}(t)}{dt} =  I_{nfall}^{\sigma} e^{-t/\tau} 
\end{equation}
where the parameter $I_{nfall}^{\sigma}$ is equal to $16 M_{\odot}Gyr^{-1}pc^{-2}$,
and this law is a function of surface gas density.
However, we want to use this law for our box, and so considering
the side of the cube ($2\cdot10^{4}pc^{2}$) as the typical surface
we obtain in each volume, multiplying the previous equation by
 this typical surface:
\begin{equation}
\frac{dM_{tot}^{Vol}(t)}{dt}= I_{nfall}^{Vol} e^{-t/\tau}    
\end{equation}
where $I_{nfall}^{Vol}$ is $320 M_{\odot}Myr^{-1}$;
$\tau$ is in both cases 1Gyr. 
We consider a threshold in the gas surface density, which means
that below this threshold the star formation is blocked, due to
the insufficient gas density in the system.
The value of this gas surface density threshold is
 $\sigma_{threshold}= 4 M_{\odot}pc^{-2}$,
 as in the homogeneous model, so with the surface we consider,
it becomes in each volume $M_{threshold}= 80\cdot 10^{3} M_{\odot}$.
The star formation rate $\psi(t)$ is defined:

\begin{equation}
\psi(t)= \nu (\frac{M_{gas}(t)}{M_{threshold}})^{1.5}
\end{equation}

where $M_{gas}(t)$ is the gas mass inside the considered box
and the parameter $\nu$ is $80 M_{\odot}Myr^{-1}$; this parameter
 is set to match the star 
formation rate of the homogenous model during the halo phase.
So, when the gas density threshold is reached for the first time, 
the mass that is transformed into stars in a timestep 
(hereafter, $M_{stars}^{new}$) is 80 $M_{\odot}$.
In this way, we note that the code can form at any timestep  stars of any mass
up to the considered maximum stellar mass (i.e. 80 $M _{\odot}$),
and the code  can form all the stellar masses up to 80 $M _{\odot}$.
So, with the considered surface (with the considered 
gas density threshold), we are also overcoming the risk of having a bias 
toward low mass stars,  when the code randomly chooses the masses of 
the  forming stars.

\begin{figure}
\begin{center}     
\includegraphics[width=0.49\textwidth]{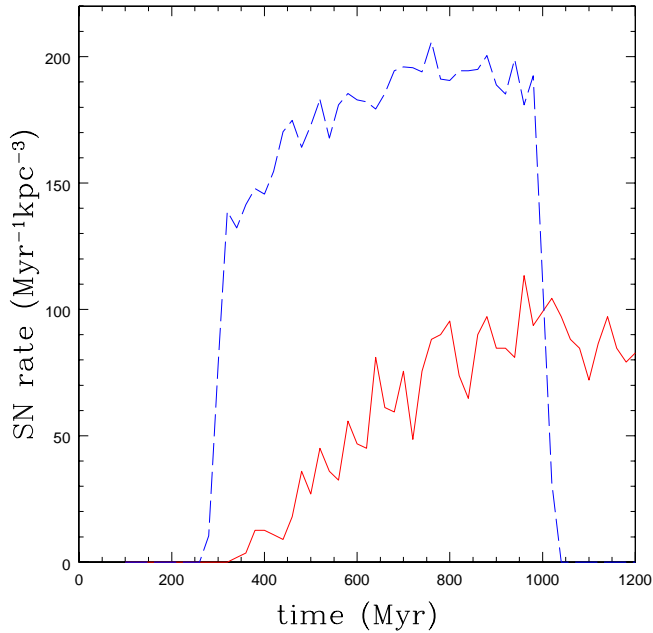} 
\caption{The SNIa rate (solid line) and SNII rate (dashed line)
in the halo. Note that the rate of SNIa is multiplied by ten.}
\label{inomo1}
\end{center}
\end{figure}

Knowing the $M_{stars}^{new}$, we assign the mass to one star with a random function,
 weighted according to the initial mass function (IMF) of Scalo (1986) in the range between 0.1 
and 80 $M_{\odot}$.
Then we extract the mass of another star and we repeat this cycle until
the total mass of newly-formed stars exceeds $M_{stars}^{new}$.
In this way, in each region, at each timestep, the $M_{stars}^{new}$ 
is the same but the total number and mass distribution
 of the stars are different, and we know the mass of each star
contained in each region, when it is born and when it will die, assuming
the stellar lifetimes of Maeder \& Meynet (1989).
We compute the chemical evolution in the following way:
at the end of its lifetime, each star
 enriches the interstellar medium
  with  newly-produced elements (see the next Sect.) 
as a function of its mass and metallicity. 
The total mass of each element is determined at the end
of the lifetime of each star, taking into account the enrichment 
due to that star and due to the already present mass of each element
locked in that star when it is born. 
The model does not take into account the pollution produced by stars 
with mass $<3M_{\odot}$ because their lifetimes exceed the 
duration of the simulation. 
The existence of SNe Ia is also taken into account, according to the prescriptions
 of Matteucci \& Greggio (1986), in the single degenerate scenario.
We consider that a fraction of the stellar masses in the
range 3-16$M_{\odot}$ would be binary systems which can produce SN Ia. 
The correct number of observed SN Ia at the present time (see Cappellaro et al. 1999)
is reproduced if this percentage is 5\% and we evaluate, 
again using a random extraction, if the system is a progenitor of
a SN Ia. We also determine the mass of the secondary through 
the distribution function:
\begin{equation}
f(\mu)=2^{1+\gamma}(1+\gamma)\mu^{\gamma}
\end{equation}
with $\gamma=2$, where $\mu$ is the fraction of the mass of the secondary star to the total mass of the system
and in this way we know when the SN Ia will explode; in fact, its lifetime is that of the
secondary star of the binary system.
In Fig. \ref{inomo1}, the SNe Ia rate is compared to the SNe II rate.
Due to the threshold in the gas density that we impose, the star formation starts only after 250 Myr and, 
at this stage, the first SNe II start to explode. With a time delay of about 30Myr, the first SNe Ia
take place. In our model we consider the single degenerate scenario and  30Myr is 
the shortest timescale for a SN Ia to explode, being the lifetime of the 
most massive progenitor leading to the formation  of  a C-O white dwarf (i.e. a star of 8$M_{\odot}$).
We stop the star formation at 1 Gyr and the SNe II rate falls abruptly to zero, correctly,
the largest lifetime of a SNII being  30Myr, whereas the SNe Ia continue to explode, the lifetime
of the progenitors of a SN Ia being as long as 10 Gyr.

We assume that the ejecta produced by the dying stars mix with the surrounding ISM only 
after these ejecta cool down significantly. The cooling timescale of the ejecta is calculated
as follows:
\begin{equation}
t_{cooling}=\frac{3kT}{n\Lambda(T,z)}
\end{equation}
where n, T, and z are the average number density, the temperature, and the metallicity in the cubic region, 
respectively. $\Lambda(T,z)$ is the cooling function, taken from Sutherland \& Dopita (1993).
We know the total gas mass and the thermal energy (assuming $10^{51}$erg of energy after each SN
explosion), therefore in each cubic region, at each timestep, we can calculate n and T.
The resulting cooling timescales are normally $\sim$ 0.1-0.2Myr
and reach at maximum 0.8 Myr in a set of 100 volumes. Since 
the cooling timescales are always smaller than the timestep of our
simulation (1 Myr),  we decided to neglect this delay time, so that the 
mixing in each simulated cubic volume is instantaneous.
The model  follows the chemical evolution of 10 elements (Si, Ca, Mg, Fe, Sr, Y, Zr, Ba, La and Eu)
in each region.
If the model is correct, our predictions will approximate the results of the homogeneous model,
having the same parameters, as the number of stars increases.  
On the other hand, our model shows the spread of chemical 
enrichment at low metallicity that can be produced by different stars
of various masses, where the number of stars is low and
the random effects in the birth of stellar masses are important.

\section{ Nucleosynthesis Prescriptions}
For the nucleosynthesis prescriptions of the Fe, Si, Ca, and Mg, 
we adopted those suggested by Fran\c cois et al. (2004)
both for single stars and SNeIa.
They started with the Woosley \& Weaver (1995) yields
 for massive stars at  the solar composition, metallicity independent;
they use the Iwamoto et al. (1999) yields for type
Ia SNe. Then, they compared the model results with observational data
with the aim of imposing constraints upon stellar yields. 
To best fit the data in the solar neighborhood, they modified 
the original yields by  Woosley \& Weaver (1995) 
and  by Iwamoto et al. (1999).
In particular, they changed Mg by increasing
in stars with masses from 11 to 20 $M_{\odot}$ and decreasing
in masses larger than 20 $M_{\odot}$.
The Mg yield has also been increased in SNe Ia.
The Si yields have increased slightly in stars above
40 $M_{\odot}$. Ca and Fe (the solar abundance case) in massive stars from
Woosley \& Weaver (1995) are the best to fit the abundance patterns 
of these elements since they do not need any changes. 
We underline that the site of production of $\alpha$-elements  
is the whole range of massive stars. For Fe, the main producers are 
SNeIa, but a fraction arises also from the whole range of massive stars.
For the nucleosynthesis prescriptions of the r-process contribution,
we used those of model 1  for Ba and Eu (see Cescutti et al. 2006),
 and the results of Cescutti et al. (2007a) for La. For Sr, Y, and Zr 
we used the results shown in Cescutti (2007b). 
We have assumed that these neutron capture elements 
are produced by r-process in massive stars, but only
 in the mass range 12-30 $M_{\odot}$.
These empirical yields have been chosen to reproduce 
the surface abundances for all these neutron capture elements 
in low-metallicity stars as well as the Sr, Y, Zr, Ba, Eu, and La solar 
abundances, taking in account the 
s-process contribution at high metallicities by means of the 
 yields of Busso et al. (2001) for La and Ba
and  those  of Travaglio et al. (2004) for Sr, Y,  and
Zr in the mass range 1.5-3$M_{\odot}$.
Eu is   considered to be a purely r-process element produced
 in the  same range of masses. 
This choice is extensively discussed 
in Cescutti et al. (2006) and the
 nucleosynthesis prescriptions of the r-process contribution
are summarized in Table \ref{rBa}. 
In this model, we do not take into account the s-process contribution 
for these elements because its enrichment timescale exceeds the duration 
of the simulation, this process taking place in low mass stars 
(1.5-3$M_{\odot}$).

\begin{table*}

\centering
\caption{The used prescriptions of the mass fractions of newly-produced elements
for neutron capture elements in massive stars (r-process).}
 \label{rBa}

\begin{minipage}{90mm}

\begin{tabular}{|c|c|c|c|c|c|c|}
\hline

$M_{star}$  & $ X_{Ba}^{new}$  & $ X_{Eu}^{new}$ & $ X_{La}^{new}$ & $ X_{Sr}^{new}$ & $ X_{Y}^{new}$ & $ X_{Zr}^{new}$\\

\hline\hline

12.   & 9.00$\cdot10^{-7}$ &  4.50$\cdot10^{-8}$ & 9.00$\cdot10^{-8}$ & 1.80$\cdot10^{-6}$& 3.60$\cdot10^{-7}$&1.80 $\cdot10^{-6}$  \\ 
15.   & 3.00$\cdot10^{-8}$ &  3.00$\cdot10^{-9}$ & 3.00$\cdot10^{-9}$ & 7.50$\cdot10^{-8}$& 2.10$\cdot10^{-8}$&1.65 $\cdot10^{-7}$  \\   
30.   & 1.00$\cdot10^{-9}$ &  5.00$\cdot10^{-10}$& 1.00$\cdot10^{-10}$& 3.25$\cdot10^{-9}$& 1.00$\cdot10^{-9}$&5.00 $\cdot10^{-9}$  \\

\hline\hline

\end{tabular}

\end{minipage}

\end{table*}

\section{Results}

\subsection {The ratios of $\alpha$-elements and neutron capture elements to Fe}

\begin{figure}
\begin{center}
\includegraphics[width=0.49\textwidth]{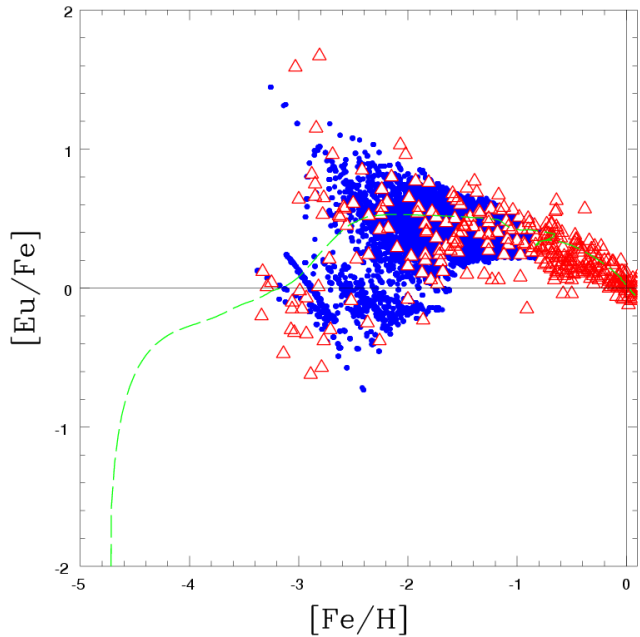}
\caption {[Eu/Fe] vs [Fe/H]. The abundances of simulated living stars at the present time
 are indicated by the  blue dots.
The red triangles are the observational data from various authors (see Table \ref{authors})
The  dashed line is the prediction of the homogeneous model (Cescutti et al. 2006, model 1).}\label{inomo2}
\end{center}
\end{figure}

\begin{figure}
\begin{center}
\includegraphics[width=0.49\textwidth]{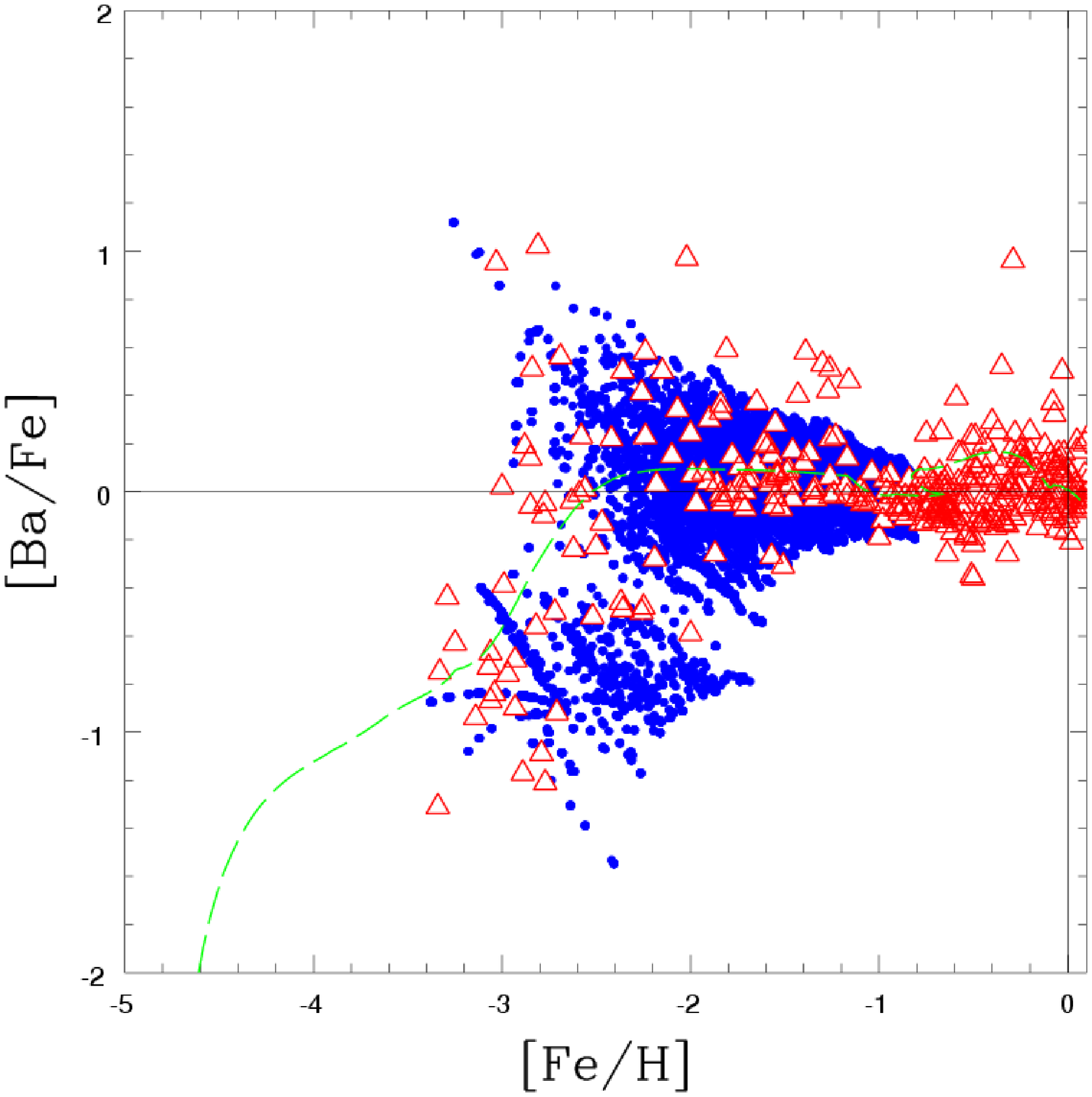}
\caption {As in Fig. 2, but for [Ba/Fe].
The authors of the observational data are listed in Table \ref{authors}.
The  dashed line is the prediction of the homogeneous model (Cescutti et al. 2006, model 1).}
\label{inomo3}
\end{center}
\end{figure}

\begin{figure}
\begin{center}
\includegraphics[width=0.49\textwidth]{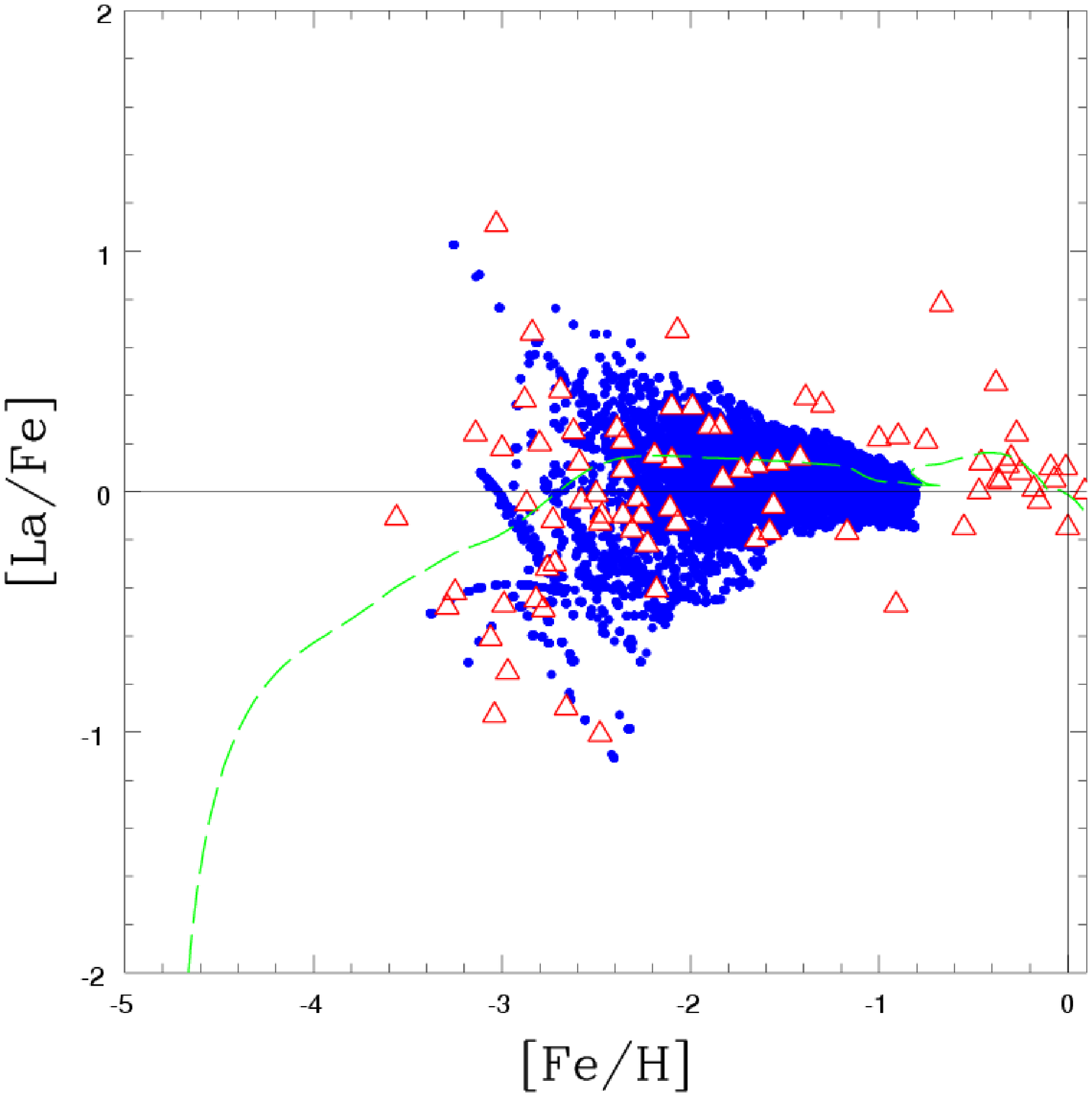}
\caption {As in Fig. 2, but for [La/Fe]. 
The authors of the observational data are listed in Table \ref{authors}.
The  dashed line is the prediction of the homogeneous model (Cescutti et al. 2007a).}\label{inomo4}
\end{center}
\end{figure}

\begin{figure}
\begin{center}
\includegraphics[width=0.49\textwidth]{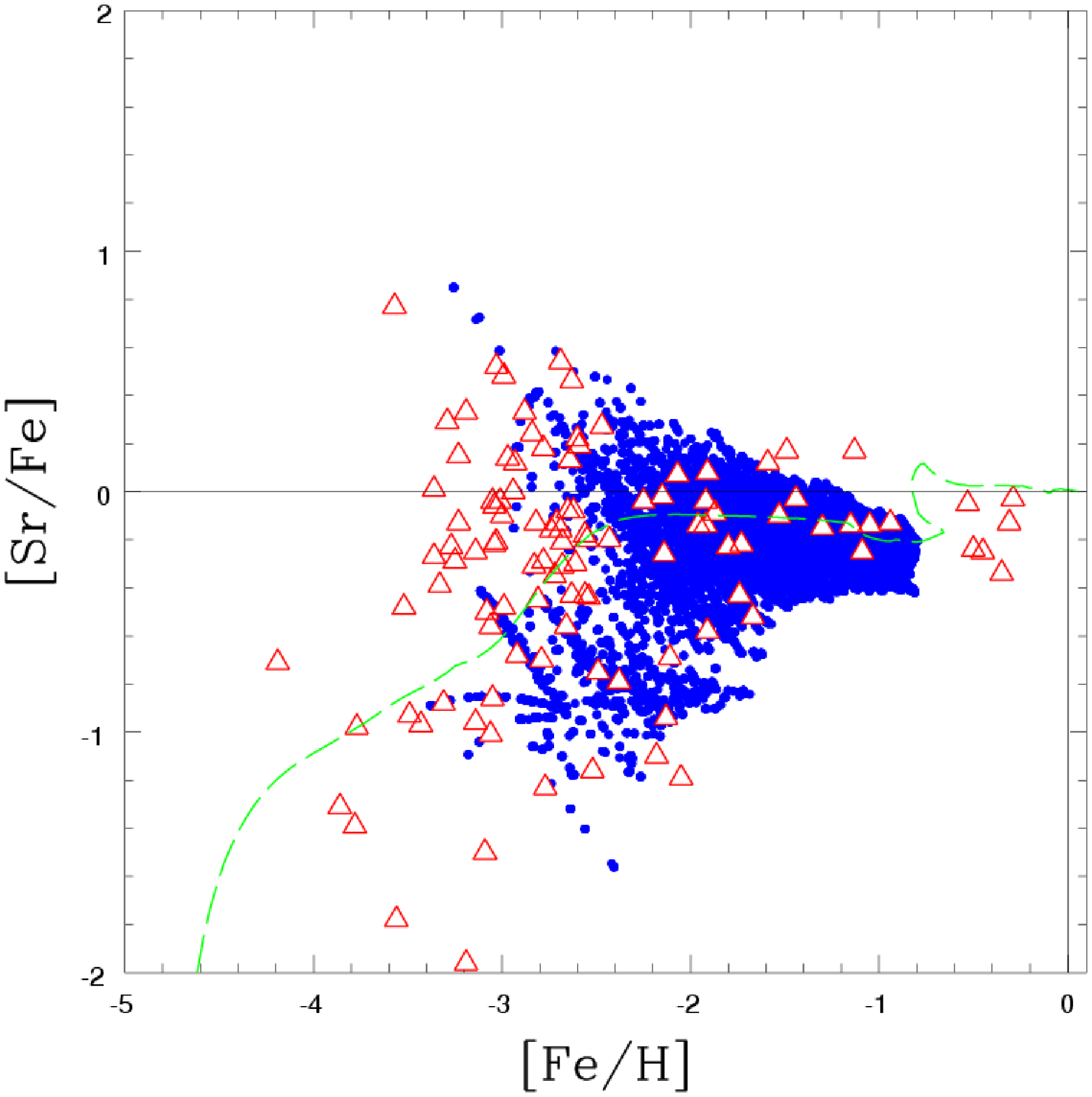}
\caption {As in Fig. 2, but for [Sr/Fe].
The authors of the observational data are listed in Table \ref{authors}. 
The  dashed line is the prediction of the homogeneous model (Cescutti 2007b).}
\label{inomo5}
\end{center}
\end{figure}

\begin{figure}
\begin{center}
\includegraphics[width=0.49\textwidth]{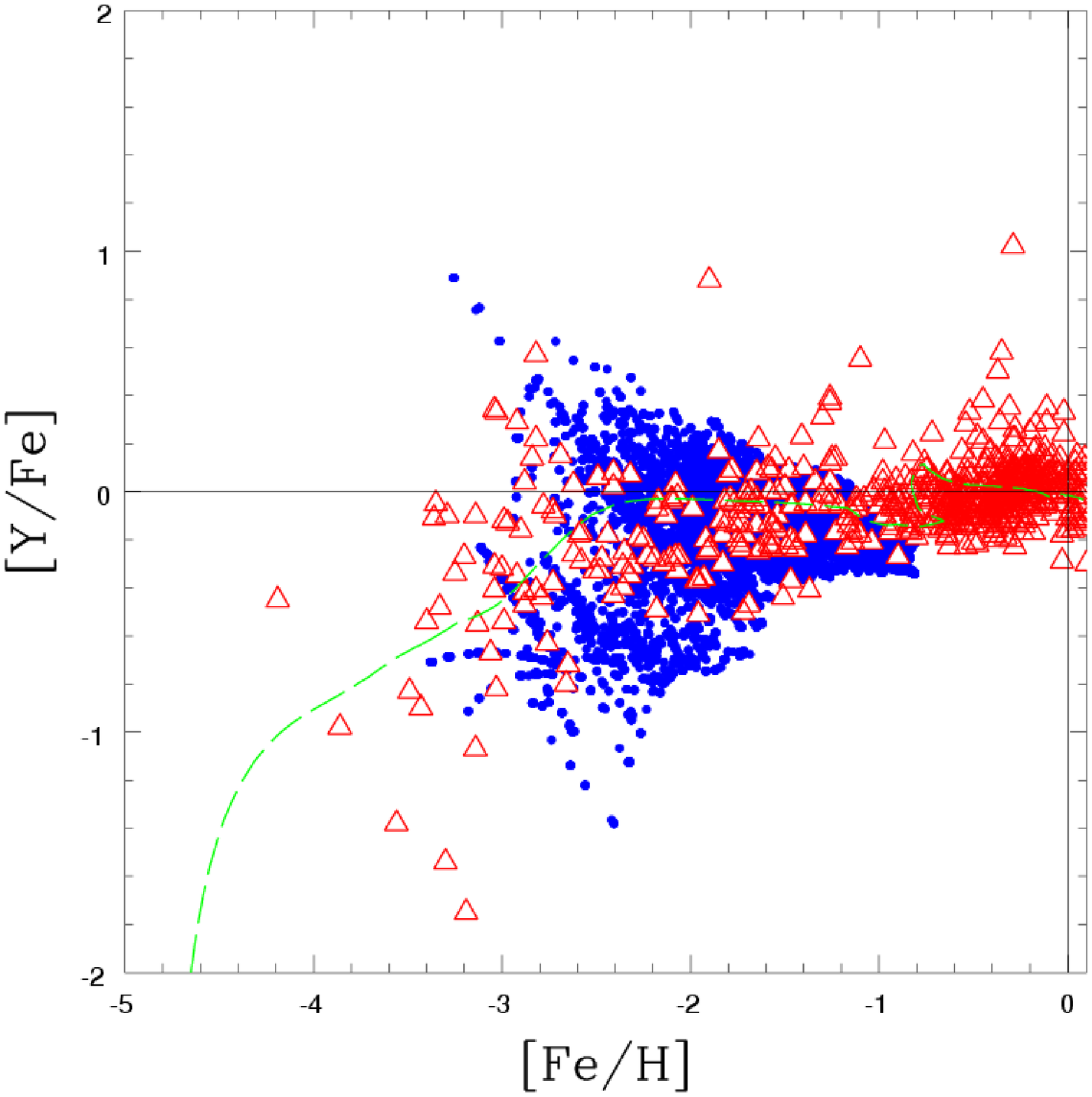}
\caption {As in Fig. 2, but for [Y/Fe]. 
The authors of the observational data are listed in Table \ref{authors}.
The  dashed line is the prediction of the homogeneous model (Cescutti 2007b).}
\label{inomo6}
\end{center}
\end{figure}

\begin{figure}
\begin{center}
\includegraphics[width=0.49\textwidth]{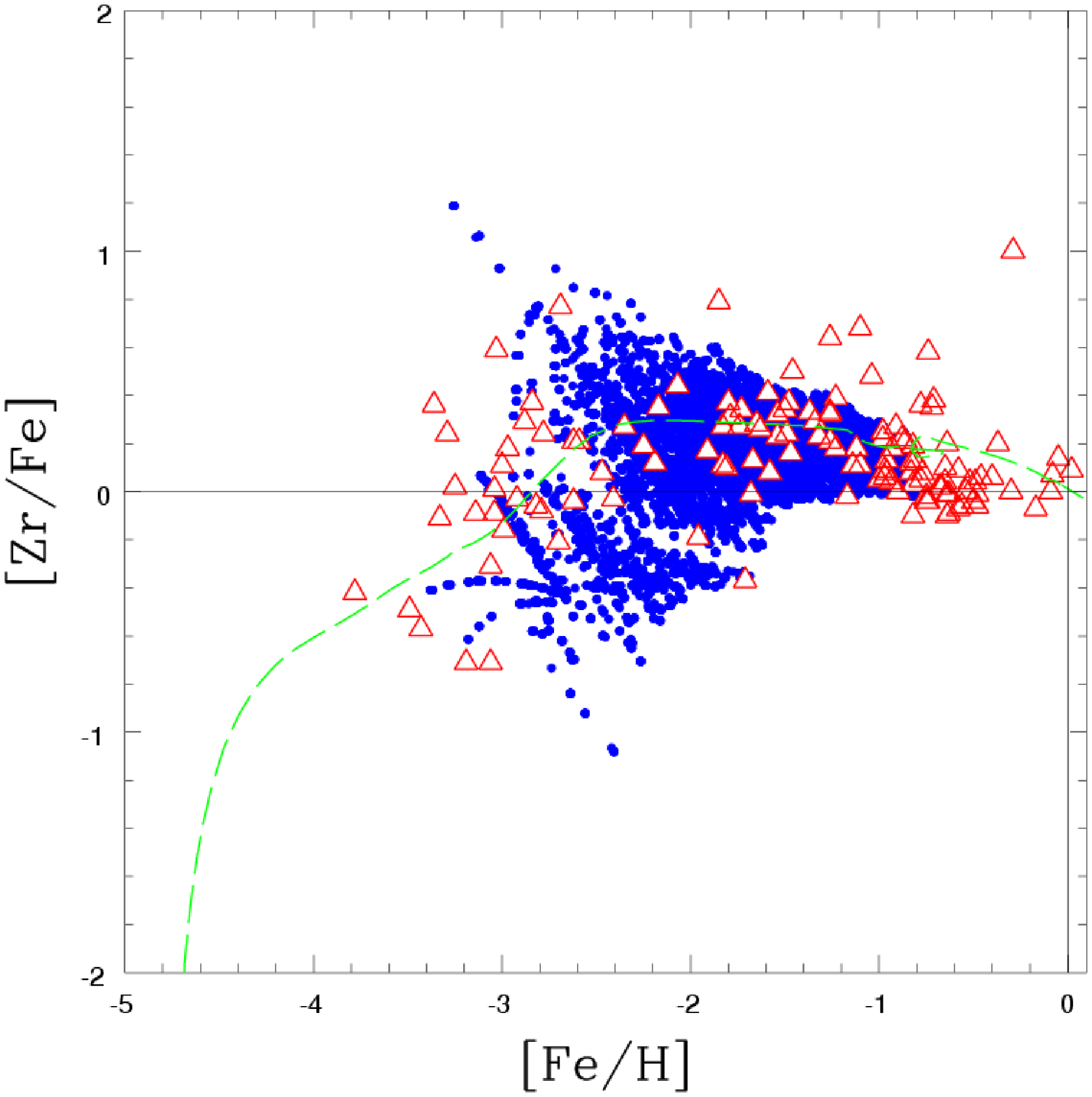}
\caption {As in Fig. 2, but for [Zr/Fe]. 
The authors of the observational data are listed in Table \ref{authors}.
The  dashed line is the prediction of the homogeneous model (Cescutti 2007b).
}\label{inomo7}
\end{center}
\end{figure}

\begin{figure}
\begin{center}              
\includegraphics[width=0.49\textwidth]{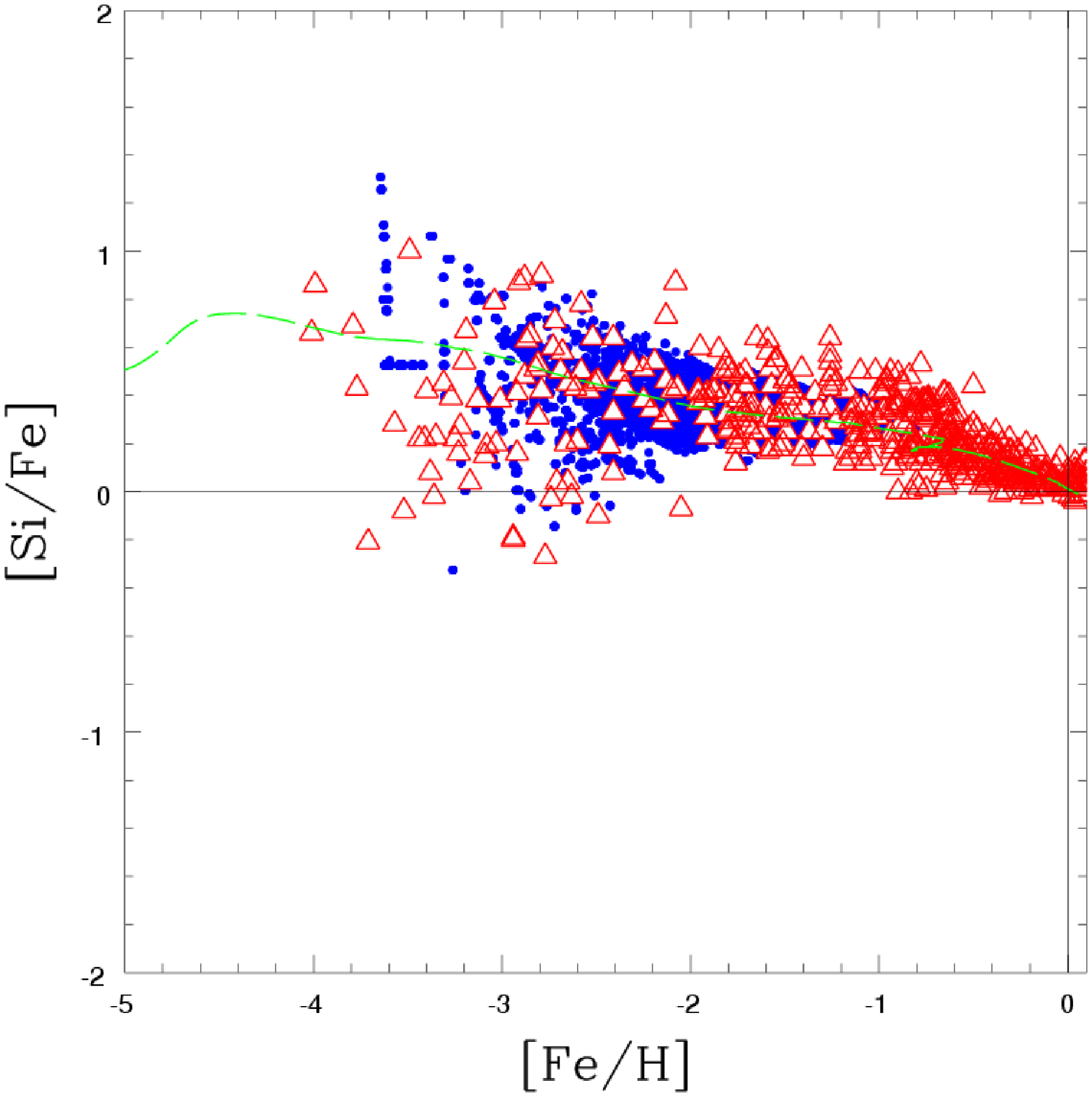} 
\caption{As in Fig. 2, but for [Si/Fe]. 
The authors of the observational data are listed in Table \ref{authors}.
The  dashed line is the prediction of the homogeneous model (Fran\c cois et al. 2004).}
\label{inomo8}
\end{center}
\end{figure}

\begin{figure}
\begin{center}              
\includegraphics[width=0.49\textwidth]{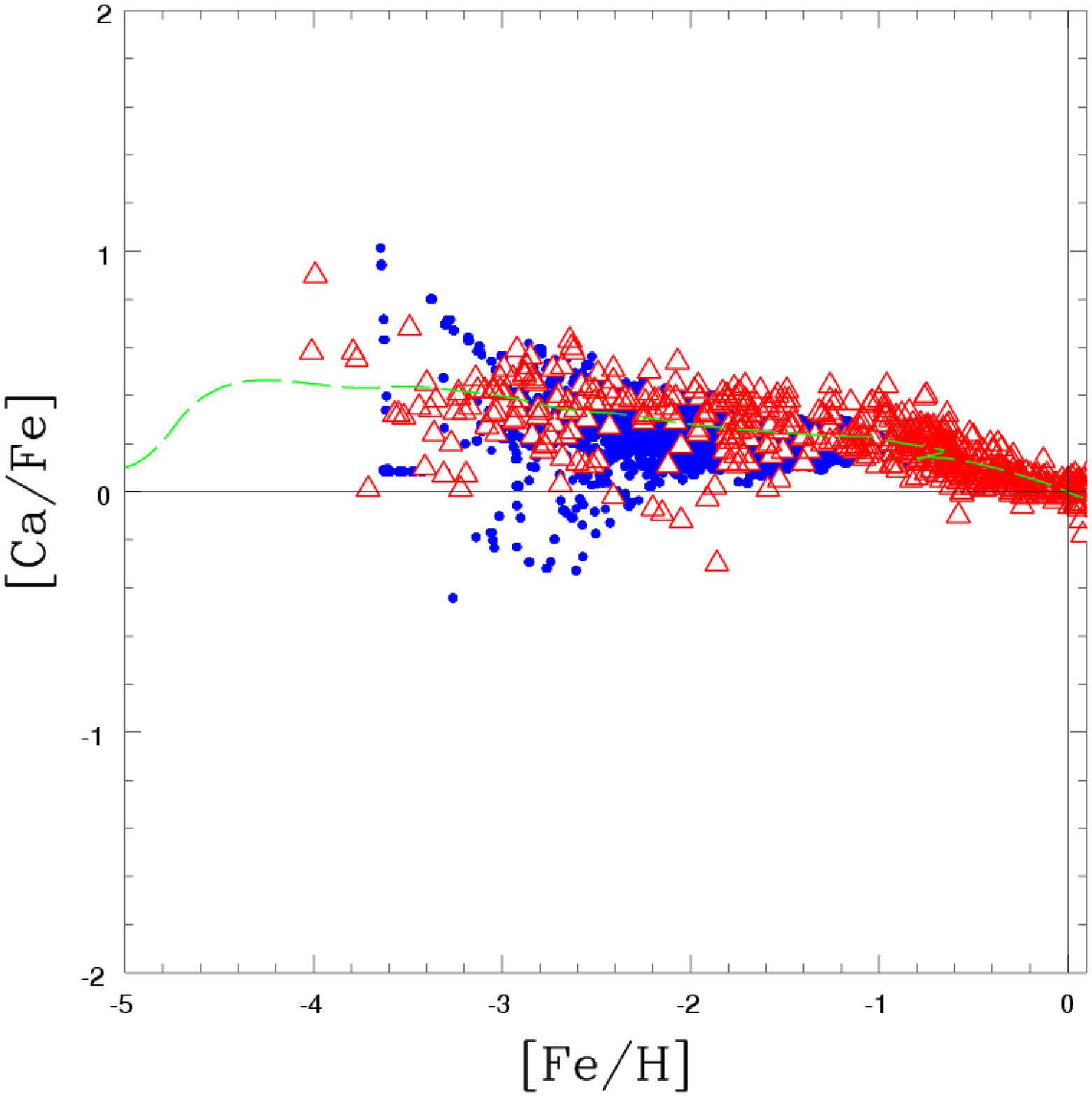} 
\caption{As in Fig. 2, but for [Ca/Fe].
The authors of the observational data are listed in Table \ref{authors}.
The  dashed line is the prediction of the homogeneous model (Fran\c cois et al. 2004).
}\label{inomo9}
\end{center}
\end{figure}

\begin{figure}
\begin{center}              
\includegraphics[width=0.49\textwidth]{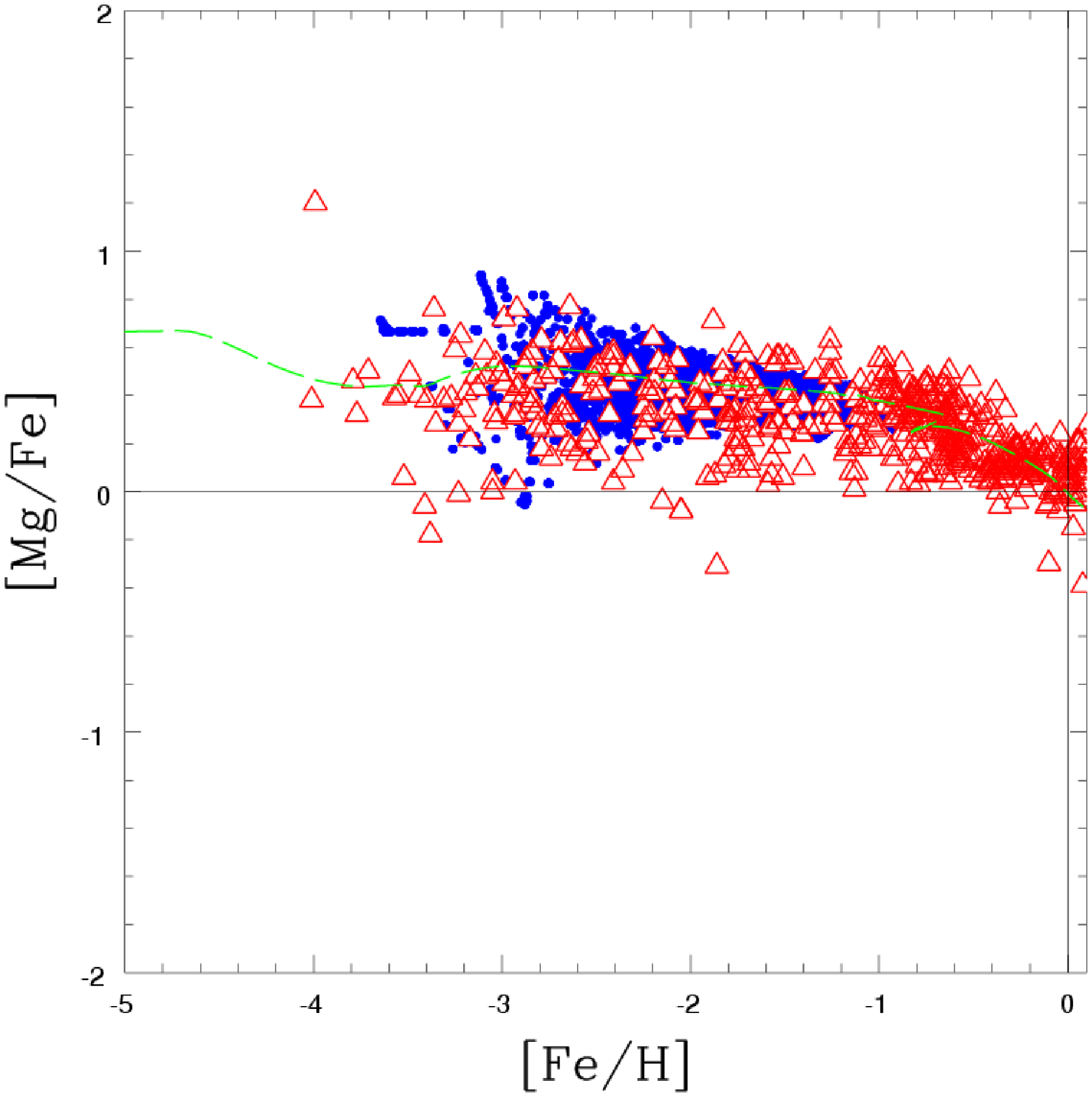} 
\caption{As in Fig. 2, but for [Mg/Fe].
The authors of the observational data are listed in Table \ref{authors}.
The  dashed line is the prediction of the homogeneous model (Fran\c cois et al. 2004).}
\label{inomo10}
\end{center}
\end{figure}

We discuss here the results of our simulations compared to the observational data and to
 the prediction of the homogeneous model of Cescutti et al. (2006).
We show the [Eu/Fe], [Ba/Fe], [La/Fe], [Sr/Fe], [Y/Fe], and [Zr/Fe] ratios 
as functions of [Fe/H] in the Figs. 
\ref{inomo2}, \ref{inomo3}, \ref{inomo4}, \ref{inomo5}, \ref{inomo6}, and \ref{inomo7}, 
respectively; for  $\alpha$-elements we show [Si/Fe], [Ca/Fe], and [Mg/Fe] as  functions of [Fe/H] 
in Figs. \ref{inomo8}, \ref{inomo9}, and  \ref{inomo10}, respectively.
In these Figs., we show the simulated living stars at the present time (blue points) together 
with the observations (red triangles).

\begin{figure}
\begin{center}              
\includegraphics[width=0.49\textwidth]{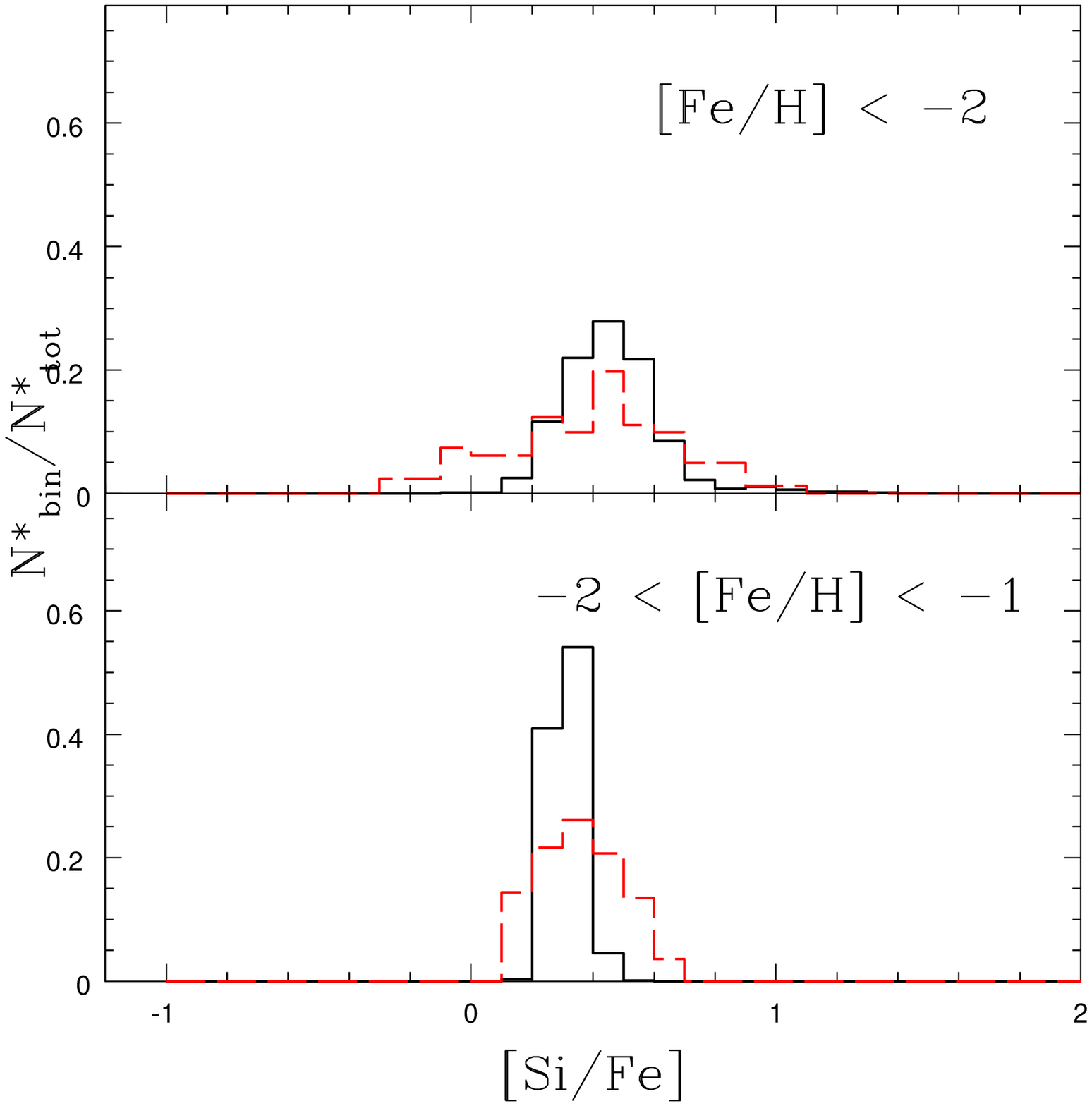} 
\caption{The relative frequency of stars at a given [Si/Fe] ratio for different enrichment phases.
In solid line the predictions of the model for the living stars at the present time,
 in dashed line the observational data.
The authors of the observational data are listed in Table \ref{authors}.
}
\label{inomo11}
\end{center}
\end{figure}

\begin{figure}
\begin{center}              
\includegraphics[width=0.49\textwidth]{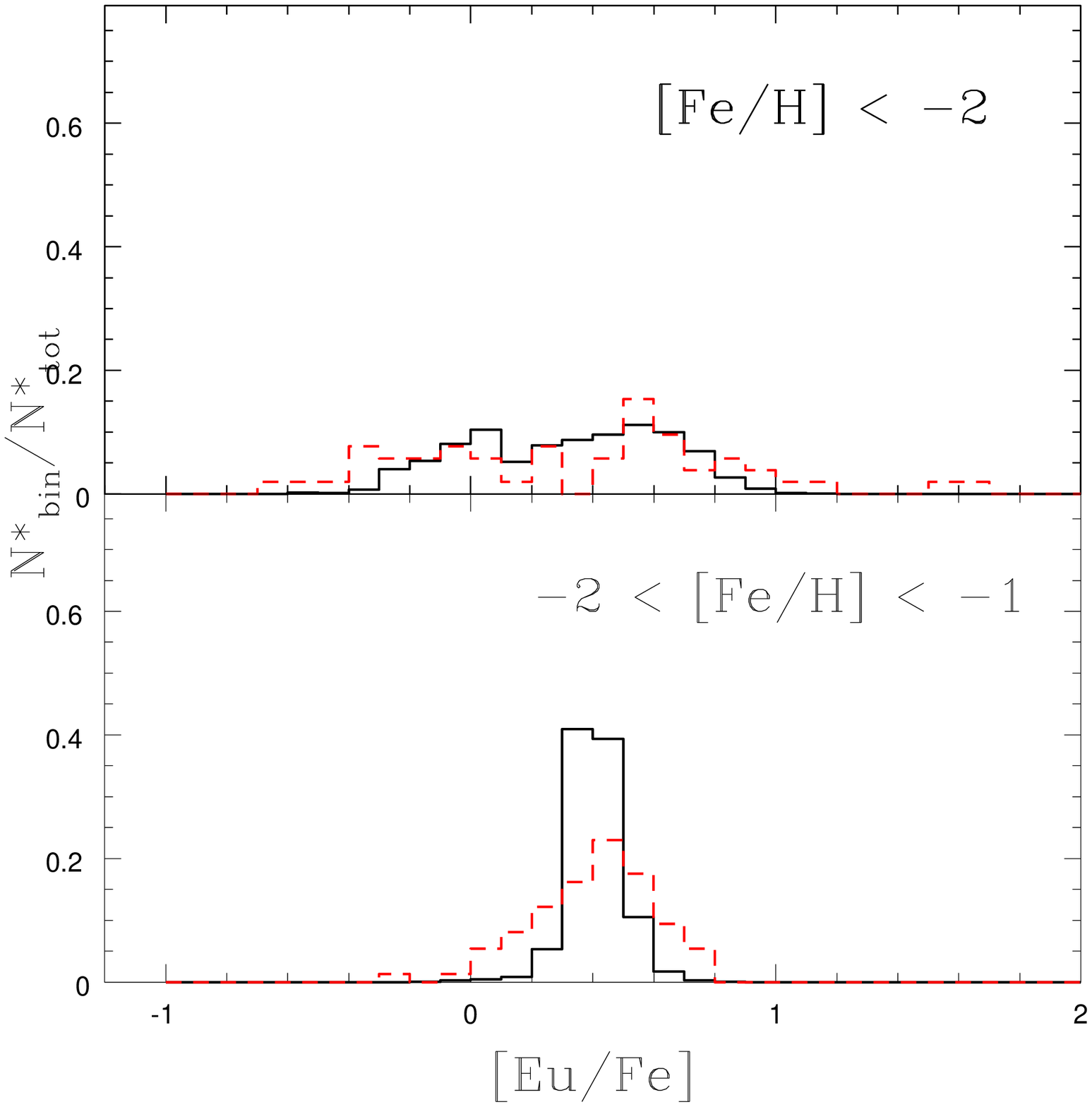} 
\caption{As in Fig. \ref{inomo11}, but for  [Eu/Fe].
The authors of the observational data are listed in Table \ref{authors}.
}
\label{inomo12}
\end{center}
\end{figure}

It is worth noting that our model reproduces the large spread observed in the  
abundances of  metal-poor stars for the neutron capture elements
 and, at the same time, the small spread for the  $\alpha$-elements.
The chemical enrichment observed in very metal-poor stars is due only 
to nucleosynthesis in massive stars.
The site of production of the
 $\alpha$-elements is  different than the one of 
neutron capture elements  (as described in the previous Sect.):
the $\alpha$-elements and Fe are produced in the whole range of massive stars (10-80$M_{\odot}$);
on the other hand, the neutron capture elements are assumed to be produced in the smaller range 
between 12 and 30 $M_{\odot}$.
Therefore, in regions biased randomly toward stars less massive than 30 $M_{\odot}$, the
 ratio of neutron capture elements over Fe is
high. The opposite happens in regions where a large fraction of the stars
 are more massive than 30 $M_{\odot}$. This fact produces, in our inhomogeneous model,
 a large spread for the rates of neutron capture elements to Fe,
 but not for the ratio of the  $\alpha$-elements to Fe, since 
the  $\alpha$-elements and Fe are produced in the same range.
  The idea of a small range of the r-elements progenitors, 
as an explanation of the large scatter, having been previously proposed 
by Ishimaru \& Wanajo (1999) and Argast et al. (2000)
(although with different mass ranges and mass dependences of the yields).
Nevertheless, Ishimaru \& Wanajo (1999) do not show  
results for  $\alpha$-elements; Argast et al. (2000) show results
for  $\alpha$-elements, but the resulting spread of their model
is too large compared to the observed one.
We have to emphasize that if we use, 
the original yields by Woosley \& Weaver,
rather than the yields by Fran\c cois et al.(2004),
 our model also shows a larger
spread than the observed spread for the Mg (but not in the case
of Ca). This is due to the strong dependence, in the original 
work of Woosley \& Weaver, of the Mg yields on stellar mass, 
compared to the smaller dependence on the progenitor mass 
in the work of Fran\c cois et al. (2004).
However, we think that it is reasonable to use the yields of Fran\c cois et al. (2004)
for Mg, since recently many  nuclear reaction data have changed,
in particular in the case of this element. More recent nucleosynthesis
yields, such as that found by Nomoto et al. (2006),
 show a smaller dependence on the progenitor mass as well.  
In the Figs. \ref{inomo11} and \ref{inomo12}, we show the relative frequency 
of stars at a given [El/Fe] ratio  for different enrichment phases.
To do that, we have used only the stars still
alive in the halo, namely those with a mass $< 0.8M_{\odot}$.
 The different enrichment phases: $[Fe/H]< -2$  and $-2<[Fe/H]<-1$
are given in the panels from top to bottom for Si and Eu.
In these Figs., the solid lines are the predictions of the model and the dashed lines are
the observational data.


At low metallicity, the model predicts  a quite narrow distribution for Si,
 whereas for Eu the distribution of the stars is broad,	in agreement with the observations.
At higher metallicities, both elements have predicted distributions
slightly narrower than the observed ones. 
Nevertheless, we emphasize that the measured ratios  are affected by
observational errors, which in general are of the order 0.1 dex and 
could be responsible for the broader distributions of the observational data.
In the low metallicity range, we noticed some discrepancies. 
For instance, the observed distribution of [Si/Fe] is slightly broader
than the predicted  distribution; therefore, a more careful analysis of nucleosynthesis yields
may be required (for instance, using metallicity dependent Si yields).
The model is not able to reproduce ultra metal-poor stars, 
 whereas stars with these metallicity are observed. In particular,
the model predicts the formation of some metal free stars at the beginning,
but then the metallicity jumps to [Fe/H]$\sim$-3.6. This may be due to an 
overly rapid rise of [Fe/H], related to the adopted infall rate.
Indeed, numerical simulations (Recchi et al. 2001) show that the mixing
timescale of the ejecta, although fast, is of the order of 10Myr, therefore larger
than the assumed timestep. If included, this delayed mixing might increase the amount
of very metal-poor stars.
Moreover, the model predicts that $\sim$ 5\% of the simulated stars are metal free.
This is the fraction of stars formed in the first 
5 Myr, in which no SN has yet exploded and enriched the ISM.
We emphasize that the star formation can be slowed down, for example, by assuming
a different infall rate or a strong outflow of gas from the halo,
 as recently adopted by Goswami \& Prantzos (2000), 
which would modulate a different star formation. This can solve 
the absence of ultra metal-poor stars and the 
too large fraction of metal free stars.
The problem of the metal free stars can be also solved by adopting
a different IMF. The results of many  works  (see 
 Larson 1998; Abel, Bryan,  \& Norman 2000; Hernandez \& Ferrara 2001;
 Nakamura \& Umemura 2001; and Mackey, Bromm, \& Hernquist 2003) predict that stars
 formed in a metal free gas  must be massive stars.
The long-living  stars, which are low mass stars, start to form when the ISM has been already 
enriched by these massive and metal free stars, the so-called Population III.
The effect of the Pop III on the global chemical evolution should be negligible  (see Matteucci \& 
Pipino 2005, Matteucci \& Calura 2005), as it involves a small total amount of recycled mass. 
In fact, we tested a model with a top-heavy IMF, for the very first period, up to 
a metallicity  $Z= 10^{-4}Z_{\odot}$;
we decided to use this value because in the literature $10^{-4}Z_{\odot}$ is considered
the metallicity of the transition to have a normal IMF (see Schneider et al. 2002).
This stage lasts only about 5-10Myr, depending on the chosen cubic region.
 This top-heavy IMF has the same shape of the 
one we normally use (Scalo et al. 1986),
but we changed the lower limit from 0.1$M_{\odot}$ to 1$M_{\odot}$. 
We found, in this way, that the chemical evolution does not change 
but we eliminated the presence of metal-free stars, which are not observed.
In fact all the stars, born during this stage will die before the present time
and the stars born after will not be  metal-free.
We emphasize that we have decided to use the same 
parameters and conditions of the homogeneous model to test the validity of the 
inhomogeneous one. It is clear from the Figs.
that this new model for a high number of star formation events well approximates
 the homogeneous model, which is shown by the dashed line in Figs. 2-10.

\subsection{The ratios [Ba/Eu] and [Ba/Y]}

Our model  predicts a spread in the abundance ratios of two elements
 if the sites of production of these two elements are different, or 
 if the ratio of the yields of these same elements presents
 large variations as a function of the stellar mass.

\begin{figure}
\begin{center}              
\includegraphics[width=0.49\textwidth]{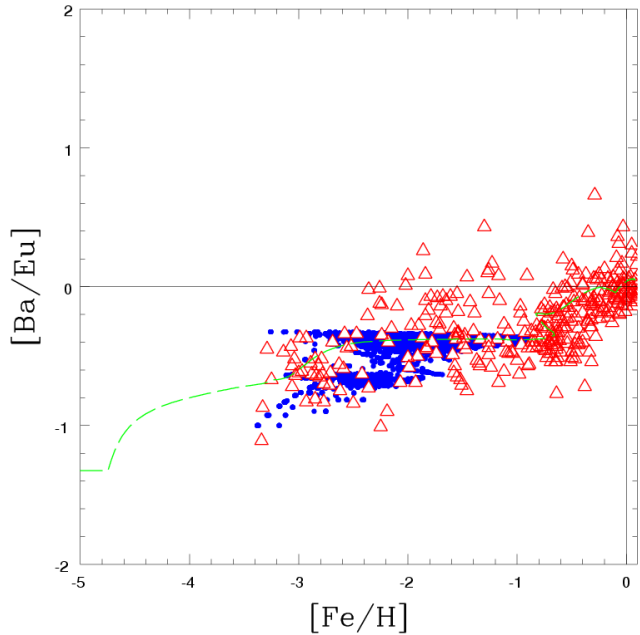} 
\caption{[Ba/Eu] vs [Fe/H]. The abundances of simulated living stars at the present time
in blue dots, the red triangles are the observational data from 
Burris et al.(2000); 
Fran\c cois et al. (2007);
Fulbright (2000, 2002); 
Gilroy et al. (1998); 
Gratton \& Sneden (1994);
Honda et al. (2004); 
Ishimaru et al. (2004); 
Johnson (2002); 
Mashonkina \& Gehren (2000, 2001); 
McWilliam et al. (1995); 
McWilliam \& Rich (1994); 
Prochaska et al. (2000); 
Ryan et al. (1991, 1996);
Stephens (1999);  		 
and Stephens \& Boesgaard (2002).
The  dashed line is the prediction of the homogeneous model (Cescutti et al. 2006, model 1).}
\label{inomo13}
\end{center}
\end{figure}

This is not the case for the yields of neutron capture elements.
In Fig. \ref{inomo13} we show the ratio [Ba/Eu] versus [Fe/H] . 
As expected, the results of the model shows a  small spread,
 as for the plot of [$\alpha$/Fe], being the  
ratio between the newly-produced Ba and the newly-produced 
Eu almost constant as a function of the stellar mass.
On the other hand, the observational data show a spread larger than predicted
in the metallicity range $-2 < [Fe/H] < -1$.
We observe that most of the data seem to have a larger overabundance of Ba
than that predicted by our model. 
Taking into account this fact, the observational spread
 could be explained either by an earlier 
production of Ba by s-process in intermediate stars from 3 to 8$M_{\odot}$ 
(that we do not take in account in our nucleosynthesis);
 by self enrichment of the observed stars due to dredge-up; or, to a
binary system with mass transfer from an AGB star, which is not shining anymore,
 to the presently observed companion star (see Aoki et al. 2006).

\begin{figure}
\begin{center}              
\includegraphics[width=0.49\textwidth]{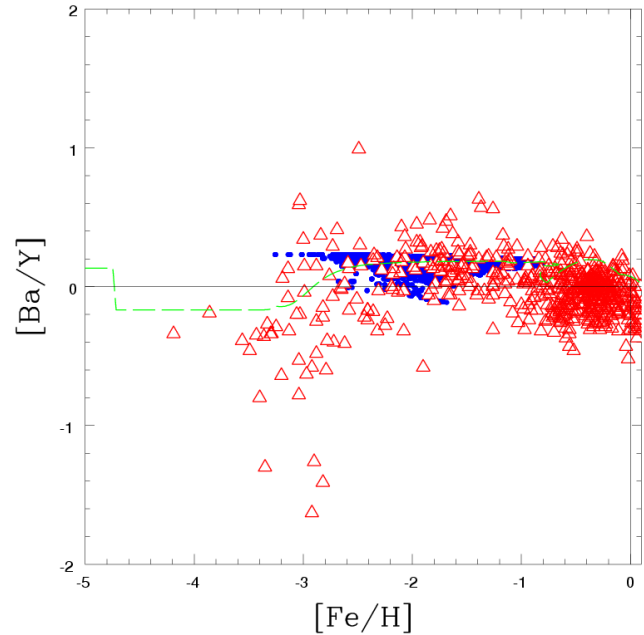} 
\caption{As in Fig. \ref{inomo13} but for [Ba/Y] vs [Fe/H].
The data for this elements are from
Burris et al.(2000); 
Fran\c cois et al. (2007);
Fulbright (2000, 2002); 
Gilroy et al. (1998); 
Gratton \& Sneden (1994);
Honda et al. (2004); 
Ishimaru et al. (2004); 
Johnson (2002); 
McWilliam et al. (1995); 
McWilliam \& Rich (1994); 
Nissen \& Schuster (1997); 
Prochaska et al. (2000); 
and Ryan et al. (1991, 1996).
The  dashed line is the prediction of the homogeneous model (Cescutti 2007b).}
\label{inomo14}
\end{center}
\end{figure}

In Fig. \ref{inomo14}, we show the ratio [Ba/Y] versus [Fe/H]. 
The results of the model relative to this ratio are
not satisfactory. The observational data show a very large scatter 
at [Fe/H] $\sim$ -3 that is not predicted by our model.
Contrary to the [Ba/Eu] ratio, for which the available data
show a moderate spread at [Fe/H] $\sim$ -3, [Ba/Y] ratio shows
a large spread. One possible explanation for this spread
 could be that the r-process yields that we use in our model are 
indicative of the mean contribution by r-process to the 
abundances of these elements. We emphasize
that these two elements are in different peaks in the solar abundances,
 both for what concern s-process (not so important at this stage)
 and for what concern r-process.
It could be that, as introduced by Otsuki et al. (2003),
the r-process is not unique but consists of different 
contributions and what we use are only the mean values.
Massive stars may produce r-process with different patterns,
probably as a function of the multiple factors that
influence r-process. 
In our inhomogeneous model, we use only one pattern for all 
the neutron capture elements. This could be the reason 
why we are not able to reproduce the spread 
for the ratios of neutron capture elements. 
Moreover, this problem should be more visible
when we compare neutron capture elements belonging to
different peaks as in the case of [Ba/Y].

\section{Conclusions}

In this paper, we tried to solve the problem of the large spread 
observed at low metallicity of the [el/Fe] ratios for neutron capture elements
and, at the same time, the smaller spread observed in [$\alpha$/Fe] ratios.
We developed a new model for the chemical evolution of the halo 
of the Milky Way. 
In this model, we coupled the birth of stars with random masses,
 with  different mass ranges for the production of
 $\alpha$-elements and neutron capture elements. 
In particular, the site of production of $\alpha$-elements is the 
whole range of the massive stars, from 10 to 80$M_{\odot}$
 whereas the mass range of production for neutron capture
 elements lies between 12 and 30$M_{\odot}$.
We showed that these two assumptions can explain the 
larger spread in the abundances of 
metal-poor stars for neutron capture elements and the smaller
spread for $\alpha$-elements. 
We note that the idea of a small range of the r-elements progenitors, 
as explanation of the large scatter, is not new, having been already proposed by 
Ishimaru \& Wanajo (1999) and Argast et al. (2000),
although the mass ranges and the mass dependences of their yields
and this work are different.
We emphasized that the parameters we used (described in Sect. \ref{model}) are 
the same as the homogeneous model (see Cescutti et al. 2006).
We adopted this point of view because these parameters have been already constrained
to give good results compared to the observational data at higher metallicities
and compared to the mean trends.
In fact, toward high metallicities ([Fe/H]$>$-2.0 dex) the model naturally gives 
results compatible to the homogeneous model.
However, this set of parameters is a starting point and the model 
still needs a better investigation of the parameter space.
In particular, concerning the early galactic stages, the model 
is not able to reproduce ultra metal-poor stars,
even though stars with this metallicity are observed.
This may be due to an overly rapid rise of [Fe/H], related to the used infall law.
To avoid these problems, a different infall law can be used or an outflow
from the halo can be included to provide a smoother increase of the metallicity.
Moreover, this new model generates an over abundance of metal free stars
if we do not impose at least a slightly different
 IMF for the first episodes of stars formation
to solve the problem of the metal free stars.

\begin{acknowledgements}
GC would like to thank Francesca Matteucci, Patrick Fran\c cois,
 Francesco Calura, Simone Recchi, and  Antonio Pipino for 
several fruitful discussions, as well as
the anonymous referee for valuable comments.
The work was supported by the Italian Ministry for the University 
and Research (MIUR) under COFIN03 prot.2003028039.
\end{acknowledgements}

\end{document}